\documentclass[prb,aps,epsfig,twocolumn,showpacs]{revtex4-1}
\usepackage{color}
\usepackage{epsfig,amssymb,amsmath,latexsym}
\usepackage{float}
\usepackage{color}
\usepackage[colorlinks=true,linktoc=page,linkcolor=blue,citecolor=magenta]{hyperref}

\newcommand\bea{\begin{eqnarray}}
\newcommand\eea{\end{eqnarray}}
\newcommand\beq{\begin{equation}}  
\newcommand\eeq{\end{equation}}

\newcommand{\non}{\nonumber}  

\newcommand\ie{{\it{i.e.}}}

\begin{document}
\textheight=23.8cm

\title{Thermal Conductance by Dirac fermions in Normal$-$Insulator$-$Superconductor junction of Silicene}
\author{Ganesh C. Paul$^{1,2}$, Surajit Sarkar$^{3}$ and Arijit Saha$^{1,2}$}
\affiliation{\mbox{$^1$} {Institute of Physics, Sachivalaya Marg, Bhubaneswar, Orissa, 751005, India} \\
\mbox{$^2$}{Homi Bhabha National Institute, Training School Complex, Anushakti Nagar, Mumbai 400085, India}\\
\mbox{$^3$}{Department of Physics, Indian Institute of Science Education and Research, Bhopal, India}
}

\date{\today}

\begin{abstract}
We theoretically study the properties of thermal conductance in a normal-insulator-superconductor junction of silicene for both
thin and thick barrier limit. We show that while thermal conductance displays the conventional exponential dependence on temperature, 
it manifests a nontrivial oscillatory dependence on the strength of the barrier region. The tunability of the thermal conductance
by an external electric field is also investigated. Moreover, we explore the effect of doping concentration on thermal conductance.
In the thin barrier limit, the period of oscillations of the thermal conductance as a function of the barrier strength comes out be 
$\pi/2$ when doping concentration in the normal silicene region is small. On the other hand, the period gradually converts to $\pi$ 
with the enhancement of the doping concentration. Such change of periodicity of the thermal response with doping can be a possible 
probe to identify the crossover from specular to retro Andreev reflection in Dirac materials. In the thick barrier limit, thermal conductance 
exhibits oscillatory behavior as a function of barrier thickness $d$ and barrier height $V_0$ while the period of oscillation becomes 
$V_0$ dependent. However, amplitude of the oscillations, unlike in tunneling conductance, gradually decays with the increase of barrier 
thickness for arbitrary height $V_0$ in the highly doped regime. We discuss experimental relevance of our results.
\end{abstract} 
\maketitle

\section{Introduction}
With the discovery of graphene~\cite{geimreview,castronetoreview} and topological insulator~\cite{sczhangreview,hasan2010colloquium}, the study 
of Dirac fermions in condensed matter systems has become one of the most active field of reseach over the last decade. The low energy band spectrum 
of these materials exhibits massless Dirac equation. Hence, relativistic electronic band structure leads to upsurge research interest in 
terms of possible application as well as fundamental physics point of view. 

In recent years, a silicon analogue of graphene, silicene~\cite{liu2011low,houssa2015silicene,ezawa2015,ytanaka2016} consisting of a monolayer 
honeycomb structure of silicon atoms, has attracted an immense amount of research interest both theoretically~\cite{ezawa2015,liu2011low} and 
experimentally~\cite{siliceneexp1,siliceneexp2,siliceneexp3,lin2012structure}. This two-dimensional (2D) material has been grown experimentally
by successful deposition of silicene sheet on silver substrate~\cite{siliceneexp1,siliceneexp2,siliceneexp3}. Also the interest in silicene
soared due to the possibility of its various future applications ranging from spintronics~\cite{RevModPhys.76.323,tsai2013gated,wang2012half,
wang2015silicene,MEzawa6}, valleytronics~\cite{ezawa2013spin,pan2014valley,kundu2016floquet,TYokoyama, saxena2015} to silicon based 
transistor~\cite{tao2015silicene} at room temperature.

Very recently, it has been reported that low energy excitations in silicene follows relativistic Dirac equation akin to 
graphene~\cite{ezawa2015,ezawa2012topological}. In fact, silicene shares almost all remarkable properties with graphene viz. hexagonal 
honeycomb structure, Dirac cones etc. However, due to large ionic radius of silicon atom, contrary to graphene, silicene does not possess 
a planar structure, rather it has a periodically buckled structure. Not only that, silicene has spin-orbit coupling 
($\sim 1.55~\rm meV$)~\cite{liu2011low} which is significantly large compared to Graphene. Consequently, a band gap appears at the Dirac points 
${\bf K}$ and ${\bf K\rq{}}$ resulting Dirac fermions to be massive. Due to the buckled structure the two sub-lattices in silicene respond 
differently to an externally applied electric field which can tune the band gap at the Dirac points~\cite{houssa2013electric,drummond2012electrically,
ezawa2012topological}. Such tunability opens up the possibility to undergo a topological phase transition from topologically non-trivial state 
to a trivial state depending on whether the applied electric field is less or more than the critical value at which the band gap closes. 
Thus a rich varity of topological phases can be realised in silicene~\cite{CCLiu2,MEzawa1,ezawa2013spin,MEzawa5,ezawa2015antiferromagnetic} 
under suitable circumstances.  

Proximity effect in Dirac materials has attracted a great deal of attention in recent times~\cite{beenakkerreview,sczhangreview}.
Very recently superconducting proximity effect in silicene has been investigated in Ref.~\onlinecite{Linder2014} in which the 
authors have theoretically studied the behavior of electrical conductance in a normal-superconductor (NS) junction of silicene. 
Up to now, no experiment has been carried out in the context of proximity effect in silicene. On the other hand, heat transport in 
Dirac systems~\cite{zhangthermal,fietethermal} and superconducting hybrid structures also has become an active field of research over the 
past decade~\cite{thermalvchandrasekhar,belzigheat,Ozaetaheat}. Thermal conductance (TC) has been investigated in graphene 
based hybrid junctions in Ref.~\onlinecite{yokoyama2008heat,wysokinski2014temperature,mohammad1,mohammad2} where due to low-energy 
relativistic nature of Dirac fermions in graphene, TC exhibits oscillatory behavior with respect to the barrier strength. Such oscillatory 
behavior of TC is in sharp contrast to that of the conventional NS junction~\cite{andreev1964thermal,bezuglyi2003heat} where TC decays with 
the barrier strength. However, study of TC in silicene based normal-insulator-superconductor (NIS) junction is still unexplored to the best 
of our knowledge. The extra tunability of the band gap by an external electric field also allows one to control the TC by the same. 
Also, TC in silicene NIS junction for both thin and thick (arbitrary barrier thickness) insulating barrier limit with different doping 
concentrations is worth to explore. 

Motivated by the above mentioned facts, in this article we study TC in silicene NIS junction for both thin and thick insulating barrier
as well as with various doping concentration in the normal silicene regime. In our analysis, we consider only the electronic part of the 
TC and neglect the phonon contribution at low temperature. We find that TC has an exponential dependance on temperature which 
is due to the s-wave symmetry of the superconductor. As the thermal transport is carried by the low-energy Dirac fermions like graphene, TC is 
shown to be oscillatory as a function of barrier strength. In moderate doped regime, where chemical potential is of the same order of band gap 
at the Dirac points, TC shows non-trivial nature due to interplay of chemical potenial, gap and temperature. TC is also controllable by the 
external electric field applied perpendicular to the silicene sheet. 
The period of oscillations of TC as a function of barrier strength depend on the doping concentration. In the thin barrier limit, the period 
changes from $\pi/2$ to $\pi$ as we go across from undoped to highly doped regime. In the thick barrier limit, oscillations 
persist in TC as a function of barrier thickness and barrier height but the period and amplitude of oscillations become functions of the barrier 
height. More strikingly, amplitude of oscillations of TC diminishes after a certain barrier thickness and height in the highly doped regime 
which is in contrast to the tunneling conductance~\cite{bhattacharjee2007theory}. 

The remainder of the paper is organised as follows. In Sec.~\ref{sec:II}, we describe our model and method. Sec.~\ref{sec:III} is devoted to 
the thin barrier regime where results are presented for three different doping concentrations. Results for the thick barrier limit are shown
in Sec.~\ref{sec:IV}. Finally, we summerize and conclude in Sec.~\ref{sec:V}.

\section{Model and Method} {\label{sec:II}}
We consider a monolayer of silicene consisting of two sublattices A and B. Two sublattice planes are separted by a distance $2l$ due to the buckled 
structure. When an electric field is applied perpendicular to the silicene sheet, a staggered potential is generated between the two sublattices 
A and B. In general tight-binding Hamiltonian for this system is given by~\cite{liu2011low,ezawa2012topological},

\begin{eqnarray}
H&=&-t\sum\limits_{<i,j>{\alpha}}\hat{c}^{\dagger}_{i{\alpha}}\,\hat{c}_{j{\alpha}}
+i {\frac{{\lambda_{SO}}}{3{\sqrt{3}}}} \sum_{<<i,j>>{\alpha}{\beta}}{\nu}_{ij}\hat{c}^{\dagger}_{i{\alpha}}{\sigma}^{z}_{{\alpha}{\beta}}\,\hat{c}_{j{\beta}}\non\\
&&-i {\frac{2}{3}} {\lambda_{R}}
\sum\limits_{<<i,j>>{\alpha}{\beta}}{\mu}_{ij}\hat{c}^{\dagger}_{i{\alpha}}{({\vec{\sigma}}\times{\hat{d}_{ij}})}^{z}_{{\alpha}{\beta}}\,\hat{c}_{j{\beta}}\non\\
&&+el\sum_{i{\alpha}}{\zeta_{i}}E^i_{z}\hat{c}^{\dagger}_{i{\alpha}}\,\hat{c}_{i{\alpha}} \,\,-{\mu}\sum_{i{\alpha
}}\hat{c}^{\dagger}_{i{\alpha}}\,\hat{c}_{i{\alpha}}\ .
\label{Hamiltonian}
\end{eqnarray}

The operator $\hat{c}^{\dagger}_{i{\alpha}}$ creates an electron at site $i$ with spin ploarization ${\alpha}$ while the operator 
$\hat{c}_{i{\alpha}}$ annihilates it. The first term describes the nearest-neighbor hopping of amplitude $t$ on honeycomb lattice, where $<i,j>$ 
denotes the nearest-neighbor sites. The second term is for the effective spin-orbit coupling (SOC) with ${\lambda_{SO}}\sim 4~\rm meV$~\cite{ezawa2012topological}, 
where ${\vec{\sigma}}=({\sigma}^x,{\sigma}^y,{\sigma}^z)$ is the pauli spin matrices and ${\nu}{_i}{_j}=({\vec{d}_i}\times{\vec{d}_j})/|{({\vec{d}_i}
\times{\vec{d}_j})}|$.
Here ${\vec{d}_i}$ and ${\vec{d}_j}$ are two nearest bonds between the next nearest-neighbors. The sum $<<i,j>>$ is over the next 
nearest-neighboring sites. The third term is the Rashba SOC of amplitude ${\lambda_{R}}$, where ${\mu}{_i}{_j}=\pm 1$ for the A(B) site. 
The fourth term represents the staggered sublattice potential, where ${\zeta}_i=\pm1$ for the A(B) sites. We consider $\lambda_{R}=0$
throughout our analysis. 
\begin{figure}[!thpb]
\centering
\includegraphics[width=0.99\linewidth]{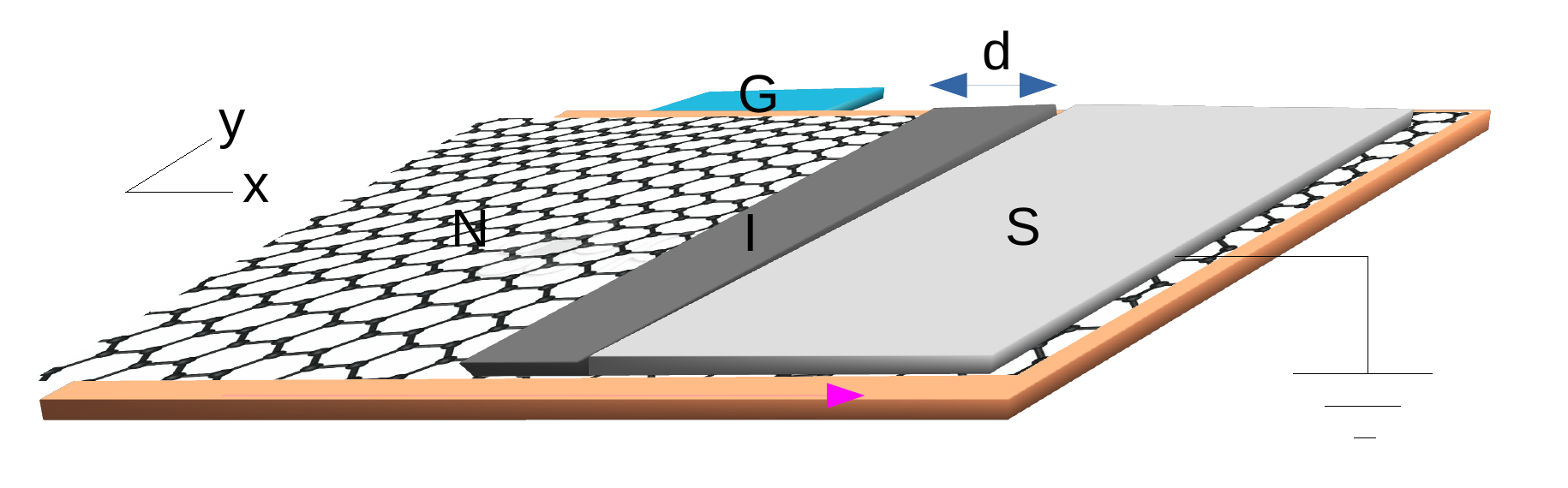}
\caption{(Color online) A schematic sketch of our silicene NIS set-up. Silicene sheet with hexagonal lattice structure is deposited on 
a substrate (orange, light grey). Here N indicates the normal region, I denotes the insulating barrier region of width $d$ (grey).  
A bulk superconducting material denoted by S (light grey) is placed in close proximity to the silicene sheet to induce superconductivity in it. 
A gate (blue, light grey) is connected to the silicene sheet to tune the chemical potential (doping) in the normal region. 
The magenta (light grey) line indicates the direction of the heat transport in response to a temperature gradient $\delta T$ between the 
normal and the superconducting side.}
\label{model}
\end{figure}
The low energy Hamiltonian of silicene can be obtained from Eq.(\ref{Hamiltonian}) near the Dirac points $k_\eta,\eta=\pm 1$ 
as~\cite{ezawa2012topological}

\begin{align}
H_\eta=\hbar v_f(\eta k_x \hat{\tau}{_x}-k_y \hat{\tau}{_y})+(elE_z-\eta\sigma\lambda{_S}{_O})\hat{\tau}{_z}-\mu \hat{1}\ .
\label{hamilt}
\end{align}

where $v_f$ is the fermi velocity of the electrons, $\mu$ is the chemical potential and $E_{z}$ is the external electric field.
$\eta=\pm 1$ corresponds to the $\bf{K}$ and $\bf{K\rq{}}$ valley.
In Eq.~(\ref{hamilt}), $\sigma$ is the spin index and $\hat{\tau}$ correspond to the Pauli matrices in the sublattice space and $\hat{1}$ 
is the $2\times2$ identity operator.

In this work we consider a normal-insulator-superconductor (NIS) set-up of silicene in $x-y$ plane as depicted in Fig.~\ref{model} with 
normal region (N) being in $x \leq -d$. The insulating region (I) with width $d$ has $-d\leq x\leq 0$ while the superconducting region (S) 
occupies $x \geq$ 0 for all $y$. The insulating region has a barrier potential which can be implemented by an external gate voltage. 
Also the chemical potential can be tuned by a gate voltage connected to the silicene sheet (see. Fig.~\ref{model}). 
Superconductivity in silicene is induced via the proximity effect of a bulk $s$-wave superconductor placed close to the silicene sheet 
in the region $x \geq 0$. 

\begin{figure}[!thpb]
\centering
\includegraphics[width=0.99\linewidth]{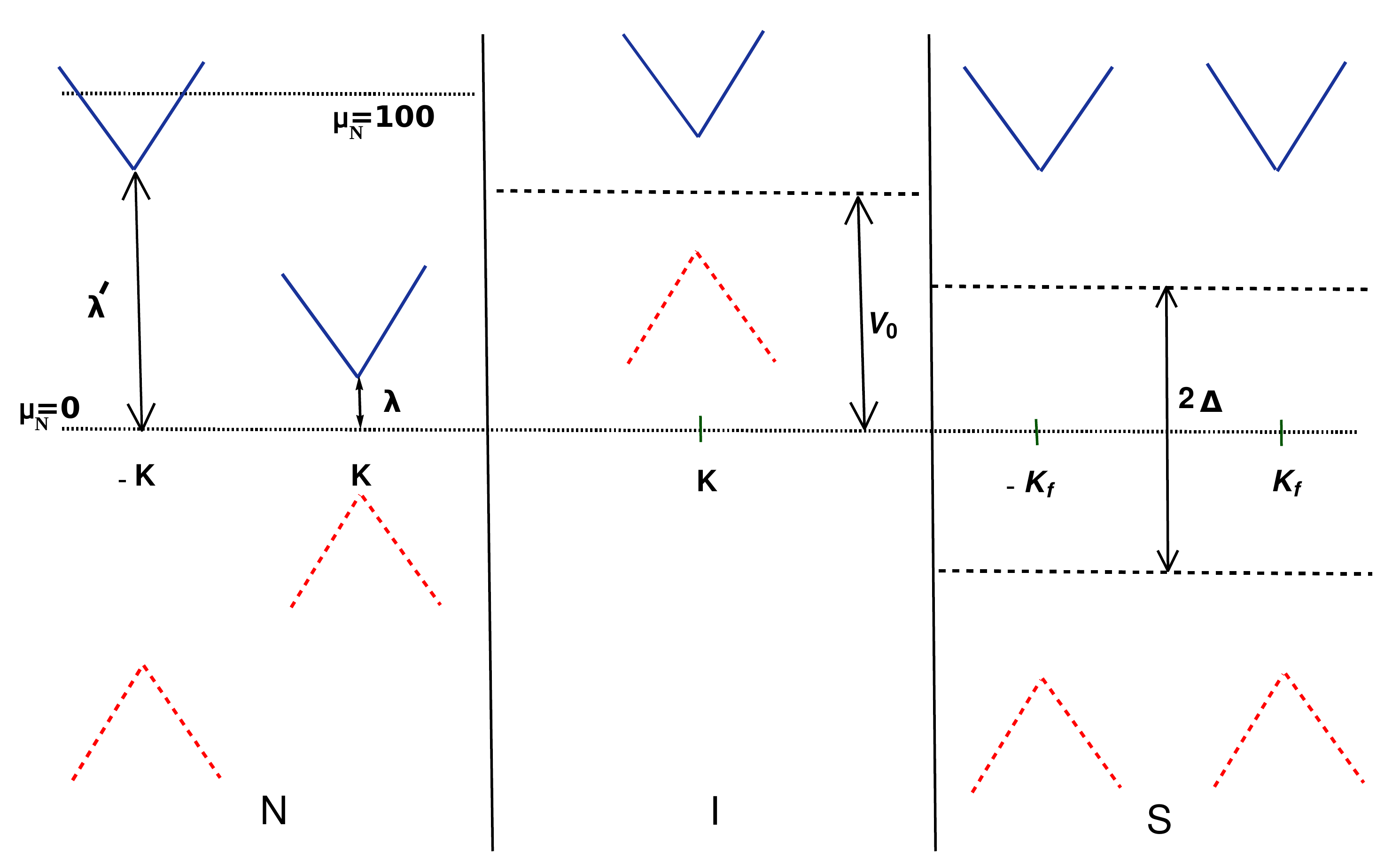}
\caption{(Color online) A schematic band diagram of our silicene NIS geometry. While in the normal (N) silicene and superconducting (S) 
silicene region both ${\bf K}$ and ${\bf K\rq{}}$ valleys are depicted, in the insulating (I) barrier region only ${\bf K}$ valley is shown
for simplicity. Blue solid line indicates conduction band while valence bands are represented by the red dashed lines. 
Dot-dashed line and dot-dot-dashed line represent $\mu_{N}=0$ and $\mu_{N}=100\Delta$ respectively.}
\label{band}
\end{figure}

Silicene NIS junction can be described by the Dirac Bogoliubov-de Gennes (DBdG) equation of the form~\cite{Linder2014}

\begin{eqnarray}
\begin{bmatrix}
\hat{H_\eta} & \Delta \hat{1}\\
\Delta^{\dagger}\hat{1}  & -\hat{H_\eta}
\end{bmatrix} {\Psi}=E{\Psi}\ .
\label{bdgHamilt}
\end{eqnarray}

where $E$ is the excitation energy, $\Delta$ is the proximity induced superconducting pairing gap and $H_\eta$ is given by Eq.(\ref{hamilt}). 
The schematic band diagram of the silicene NIS set-up is shown in Fig.~\ref{band}. In silicene, the pairing occurs
between $\eta=1$, $\sigma=1$ and $\eta=-1$, $\sigma=-1$ as well as $\eta=1$, $\sigma=-1$ and $\eta=-1$, $\sigma=1$ for a
$s$-wave superconductor.

Solving Eq.(\ref{bdgHamilt}) we find the wave functions in three different regions. The wave functions for the electrons and holes moving 
in $\pm x$ direction in normal silicene region reads

\begin{eqnarray}
\psi_N^e{^\pm}= \frac{1}{A}
\begin{bmatrix}
 \frac{\pm{\eta}k^e_1e^{\pm}{^i}{^\eta}{^{\alpha_e}}}{\tau^e_1}\\
 1\\
 0\\
 0
\end{bmatrix}
\exp[i({\pm}k^e_{1_{x}}x+k^e_{1_{y}}y)] \ , \nonumber \\
\psi_N^h{^\pm}= \frac{1}{B}
\begin{bmatrix}
 0\\
 0\\
 \frac{\mp{\eta}k^h_1e^{\pm}{^i}{^\eta}{^{\alpha_h}}}{\tau^h_1}\\
 1
\end{bmatrix}
\exp[i({\pm}k^h_{1_{x}}x+k^h_{1_{y}}y)]\ .
\label{wfn}
\end{eqnarray}

where the normalization factors are given by $A={\sqrt{\frac{2(E+{\mu_N})}{\tau^e_1}}}$,\,\,$B={\sqrt{\frac{2(E-{\mu_N})}{\tau^h_1}}}$ and

\bea
k^{e(h)}_{1}&=&\sqrt{\Big(k^{e(h)}_{1_{x}}\Big)^2 + \Big(k^{e(h)}_{1_{y}}\Big)^2}\ ,
\eea

\begin{align}
k^{e(h)}_{1_{x}}=\sqrt{(E{\pm}{\mu{_N}})^2 - (elE{_z} - {\eta}{\sigma}{\lambda}{_S}{_O})^2 - \Big(k^{e(h)}_{1_{y}}\Big)^2}\ .
\end{align}

\bea
{\tau^{e(h)}_1}&=&E{\pm}{\mu_N}{\mp}(elE{_z}-{\eta}{\sigma}{\lambda}{_S}{_O})\ .
\eea

Here $\mu_{N}$ is the chemical potential in the normal silicene region. 

Due to translational invariance in the $y$-direction, corresponding momentum $k^{e(h)}_{1_{y}}$ is conserved. The angle of incidence 
${\alpha_e}$ and the Andreev reflection (AR) angle ${\alpha_h}$ are related via the relation 

\bea
k^h_{1}\sin({\alpha{_h}})=k^e_{1}\sin({\alpha{_e}})\ .
\eea

For the rest of the paper, we have denoted the band gap $(elE_z-\lambda_{SO})/\Delta$ at ${\bf K}$ valley by $\lambda$ and the gap
$(elE_z+\lambda_{SO})/\Delta$ at ${\bf K\rq{}}$ valley by $\lambda\rq{}$. In the Insulating region wave functions can be found 
from normal region wave functions (Eq.(\ref{wfn})) by replacing $\mu{_N}\rightarrow \mu{_N}-V_0$.

In the Superconducting region the wave functions of DBdG quasiparticles are given by,
\begin{eqnarray}
\psi_S^e&=&
\frac{1}{\sqrt{2}}
\begin{bmatrix}
u_{1}\\
{\eta}u_{1}e^i{^\eta}{^{\theta_e}}\\
u_{2}\\
{\eta}u_{2}e^i{^\eta}{^{\theta_e}}
\end{bmatrix}
\exp[(i{\mu}{_S}-{\kappa})x+iq^e_{y} y]\ , \notag\\
\psi_S^h&=&\frac{1}{\sqrt{2}}
\begin{bmatrix}
u_{2}\\
-{\eta}u_{2}e^-{^i}{^\eta}{^{\theta_h}}\\
u_{1}\\
-{\eta}u_{1}e^-{^i}{^\eta}{^{\theta_e}}
\end{bmatrix}
\exp[(-i{\mu}{_S}-{\kappa})x+iq^h_{y} y]\ .
\label{wfsc}
\end{eqnarray}

Here, $u_{{1(2)}}={\Big[{\frac{1}{2}}\,{\pm}\,{\frac{\sqrt{E{^2}-{\Delta}{^2}}}{2E}}\Big]^{\frac{1}{2}}}$ and \,${\kappa}={\sqrt{{\Delta}{^2}-E^2}}$.
The transmission angles for electron-like and hole-like quasi-particles are given by,
\bea
q{^{\alpha}}\sin{\theta}_{\alpha}=k^e_1 \sin{\alpha}_{e}\ .
\label{sc1}
\eea
for ${\alpha}=e,h.$ The quasiparticle momentums can be written as
\bea
q^{e(h)}={\mu}{_S}\,{\pm}\,{\sqrt{E{^2}-{\Delta}{^2}}}\ .
\label{sc2}
\eea
where $\mu_S=\mu_N + U_0$ and $U_0$ is the gate potential applied in the superconducting region to tune the Fermi surface mismatch.
The requirement for the mean-field treatment of superconductivity is that $\mu_{S}\gg\Delta$~\cite{beenakker1,beenakkerreview}.

We consider electrons with energy $E$ incident at the interface of our NIS junction of a silicene sheet. Considering both normal reflection and 
Andreev reflection from the interface, we can write the wave functions in three different regions of the junction as

\begin{eqnarray}
\Psi_N&=&\psi_N^e{^+}+r\psi_N^e{^-}+r{_A}\psi_N^h{^-}\ ,\non\\
\Psi_I&=&p\psi_I^e{^+}+q\psi_I^e{^-}+m\psi_I^h{^+}+n\psi_I^h{^-}\ , \non\\
\Psi_S&=&t{_e}\psi_S^e+t{_h}\psi_S^h\ .
\label{bc}
\end{eqnarray}

where $r$ and $r{_A}$ are the amplitudes of normal reflection and Andreev reflection in the N region respectively. $t{_e}$ and $t{_h}$ denote 
the amplitudes of transmitted electron like and hole like quasiparticles in the S region. Using boundary conditions at the two interfaces, 
we can write

\begin{eqnarray}
\Psi_N|{_{x=-d}}=\Psi_I|{_{x=-d}},\,\,\,\, \Psi_I|{_{x=0}}=\Psi_S|{_{x=0}}\ .
\label{bc1}
\end{eqnarray}

From these equations we can find the reflection and AR amplitudes $r$ and $r_{A}$, required for evaluating the electronic contribution of TC. 
For the NIS junction, normalized thermal conductance $\kappa$ is given by~\cite{bardas1995peltier,yokoyama2008heat}

\begin{eqnarray}
\kappa&=&\int_{0}^{\infty}\int_{-\frac{\pi}{2}}^{\frac{\pi}{2}} dE d{\alpha}_{e}
\Bigg[1-R_e-R_h\frac{\cos({\alpha}_h)}{\cos({\alpha}_e)}\Bigg]\non\\
&&\cos({\alpha}_e)\,\Bigg[\frac{E^2}{4T^2\cosh^2 ({\frac{E}{2T}})}\Bigg]\ .
\label{kappa}
\end{eqnarray}

Here, $R_e$ and $R_h$ are reflection and AR probability respectively. From current conservation, we obtain~\cite{Linder2014}
\begin{eqnarray}
R_e &=&|r|^2 \ ,\nonumber\\
R_h &=&\frac{k^{h}_{1_{x}}}{k^{e}_{1_{x}}} \Bigg[\frac{2(E+\mu_N)(E-\mu_N-\lambda)}{|\eta k^{h}_{1_{x}}-i k^{e}_{1_{y}}|^2+(E-\mu_N-\lambda)^2}\Bigg] 
|r_A|^2\ .
\end{eqnarray}

\section{Thin barrier}{\label{sec:III}}
In this section, we present our numerical results based on Eq.(\ref{kappa}) for the thin barrier limit. In this particular limit of 
insulating barrier, we consider $d\rightarrow0$ and $V_{0}\rightarrow{\infty}$, such that $k_I^{e}d,k_I^{h}d\rightarrow{\chi}$ where 
$k_I^e, k_I^h$ are the electron and hole momentum inside the insulating barrier respectively. ${\chi}$ is defined as the barrier strength 
of the insulating region. Such limit has been considered before in Ref.~\onlinecite{subhro} for the analysis of tunneling conductance
in graphene. 

We take $U_0$ to be very large compared to the superconducting pairing potential $\Delta$. For simplicity, we consider $\theta_e=0$ and 
$\theta_h=0$ in Eq.(\ref{sc1}) and Eq.(\ref{sc2}).
Due to significant chemical potential imbalance between the normal and superconducting sides, there is a large mismatch of Fermi wavelengths 
in these two sides resulting in interesting behavior in TC. 

Before proceeding to present our numerical results, here we illustrate whether both valleys contribute to TC at all doping conentrations or not. 
For silicene, the band gap at $\bf{K\rq{}}$ valley satisfies $\lambda\rq{}\,\gg\,\mu_{N}/\Delta$ for the undoped and moderately doped regime. 
Consequently, $\bf{K\rq{}}$ valley does not contribute to TC in these two regimes. Nevertheless, in highly doped regime, $\mu_{N}\sim 100\Delta$ 
which is much larger than both the band gaps $\lambda$ and $\lambda^{\prime}$ at $\bf{K}$ and $\bf{K\rq{}}$ valley respectively (see Fig.~\ref{band}). 
Hence, we consider contribution for both the valleys while calculating TC for the highly doped regime. Therefore, we can write 
$\kappa\,=\,{\kappa}_{\bf{K}}+{\kappa}_{\bf{K\rq{}}}$ in this case. On the other hand, $\kappa\,=\,\kappa_{\bf{K}}$ for the undoped 
and moderately doped case. 

\begin{figure}[!thpb]
\centering
\includegraphics[width=0.99\linewidth]{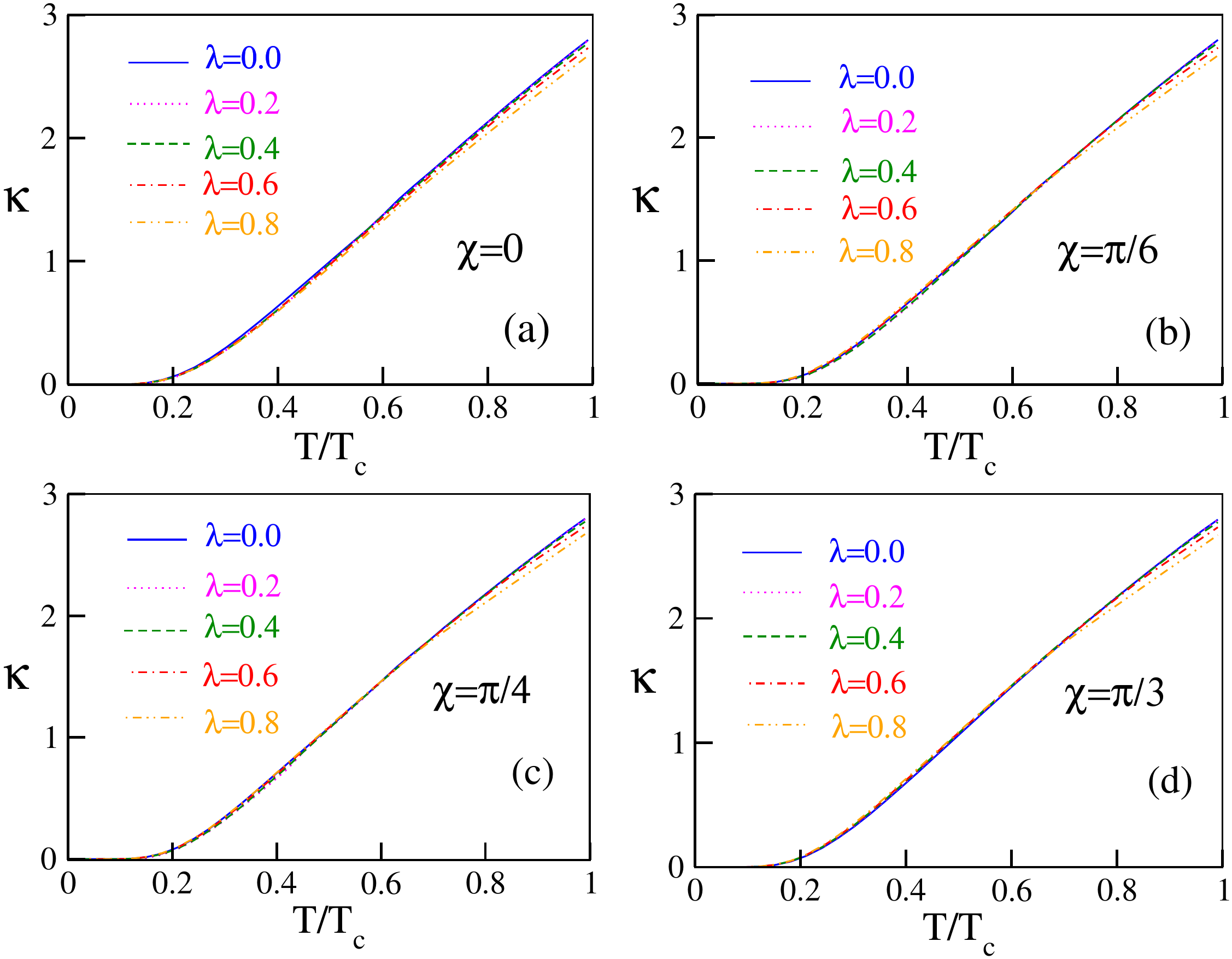}
\caption{(Color online) Thermal conductance is shown as a function of temperature $T/T_C$ with $U=100\Delta$ and $\lambda$ ranging 
from 0 to 0.8 for the undoped regime ($\mu_{N}=0$). 
}
\label{KTMu0}
\end{figure}

\begin{figure}[!thpb]
\centering
\includegraphics[width=1.0\linewidth]{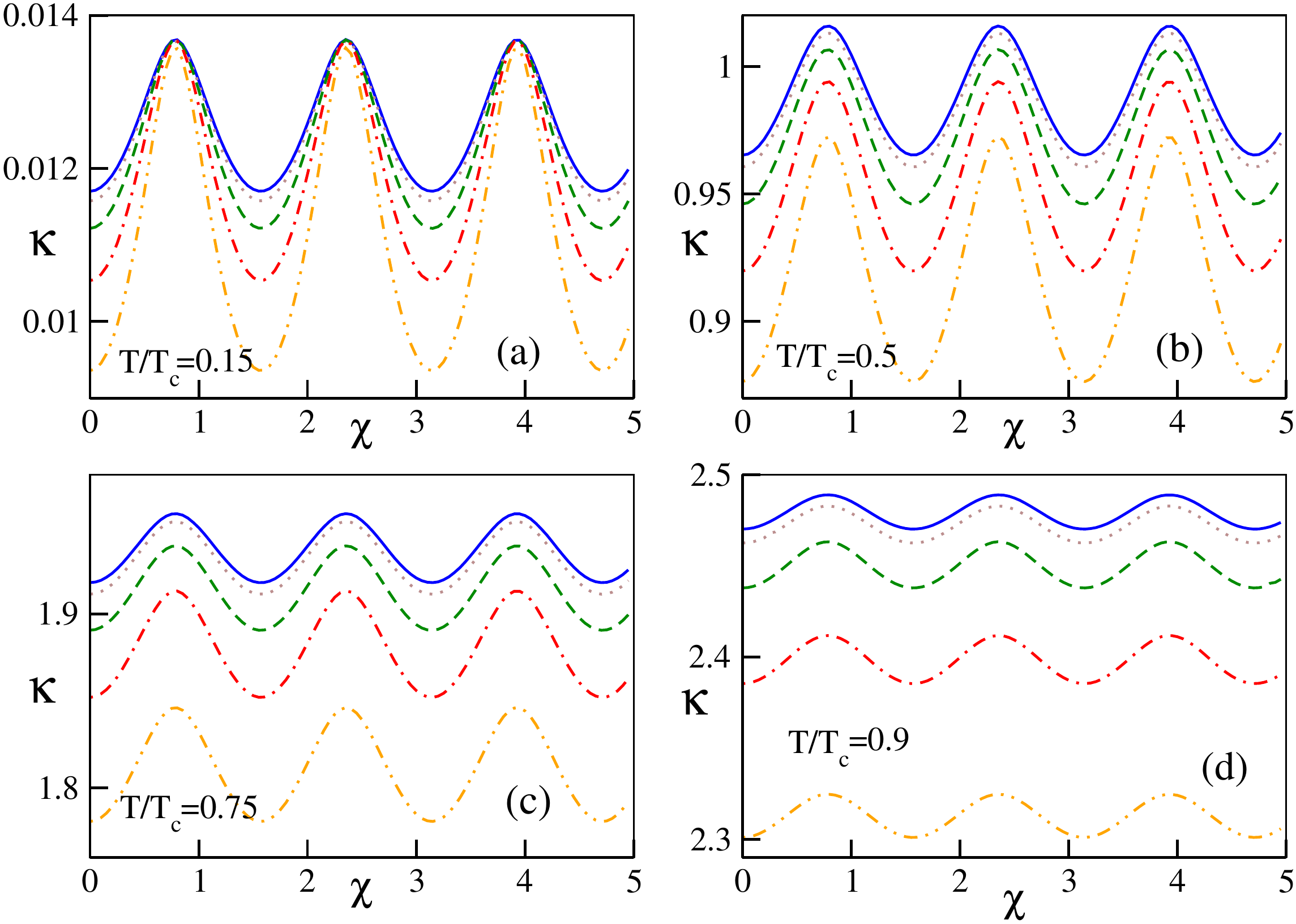}
\caption{(Color online) Thermal conductance is depicted as a function of the barrier strength $\chi$ with $U=100\Delta$ and $\lambda$ 
ranging from 0 to 0.8 for the undoped regime ($\mu_{N}=0$). Blue (solid), magenta (dotted), green (dashed), 
red (dash-dotted) and orange (dash-dot-dotted) curves indicate $\lambda$ values 0.0, 0.2, 0.4, 0.6 and 0.8 respectively.}
\label{KChiMu0}
\end{figure}

\subsection{Undoped regime ($\mu_{N}=0$)}
In this subsection we present our results when the normal region of the silicene sheet is undoped \ie~$\mu_{N}=0$. In Fig.~\ref{KTMu0}((a)-(d))
we show the behavior of TC as a function of $T/T_c$ for $\lambda$ ranging from $0$ to $0.8$. In silicene $\lambda$ can be tuned by just the 
external electric field $E_{z}$. We choose various barrier strengths. 
Here, $T_c$ is the transition temperature of the superconducting silicene. The exponential fall of TC ($\kappa$) when the 
temperature is below the transition temperature $T_c$ results because of spherical symmetry of the $s$-wave superconductor~\cite{yokoyama2008heat}. 
This behavior is similar to that of conventional nomal metal-superconductor junction~\cite{bezuglyi2003heat}. 
As we increase $\lambda$ by suitably adjusting the perpendicular electric field $E_{z}$, 
TC decreases monotonically. As $\lambda$ \ie~band gap increases, the available propagating states through which thermal transport takes place 
reduces and as a consequence TC decreases monotonically with $\lambda$. However, as carriers with all energies contribute to the thermal 
transport, quantitative value of $\kappa$ is hardly affected by change of band gap at Dirac points which is less than the induced 
superconducting gap in magnitude. This we can see from formula of $\kappa$ (see Eq.(\ref{kappa})).

In Fig.~\ref{KChiMu0}((a)-(d)) we demonstrate the bahavior of TC with respect to the barrier strength $\chi$. We choose different temperatures
below the transition temperature $T_{c}$. TC exhibits a periodic behavior with periodicity $\pi/2$ as shown in Fig.~\ref{KChiMu0}.
Such periodic behavior of TC is entirely different from conventional NS junction where TC always decays with the barrier strength. 
This periodic behavior is also the manifestation of Dirac fermions in silicene. 
When temperature $T$ is very small compared to $T_c$, the quantitative value of $\kappa$ is vanishingly small which can also be seen from 
Fig.~\ref{KTMu0} focusing at small $T/T_c$ region. Also, the $\pi/2$ periodicity of TC is independent of $T/T_c$ value. 
Moreover, for the $\mu_{N}=0$ regime, the major contribution in TC originates from the specular Andreev reflection (SAR)~\cite{beenakker1} 
process due to the pecularity of 2D Dirac systems. Effect of $\lambda$ is more prominent near the transition temperature $T_c$ because 
superconducting gap decreases as $T\rightarrow T_c$ resulting the band gap in the normal region to overcome the superconducting pairing gap. 
As a result normal reflection probability enhances rusulting in reduction in $\kappa$. Note that, the maxima of the peaks of $\kappa$ for 
different $\lambda$ are same for $T\,=\,0.15 T_c$ which is unique behavior at very low temperature ($T\ll T_c$). On the contrary, peak heights 
of $\kappa$ gets reduced due to the evanescent modes as long as $T$ approaches $T_c$. 

The oscillatory behavior of the TC can be explained as follows. Nonrelativistic free electrons with energy $E$ incident on a potential barrier 
with height $V_0$ are described by an exponentially decaying (non-oscillatory) wave function inside the barrier region if $E < V_0$, 
since the dispersion relation is $k\sim \sqrt{E-V_0}$. On the contrary, relativistic free electrons satisfies a dispersion $k\sim (E-V_0)$, 
consequesntly corresponding wave functions do not decay inside the barrier region~\cite{lindersudbo}. Instead, the transmittance of the junction 
displays an oscillatory behavior as a function of the strength of the barrier. Hence, the undamped oscillatory behavior of TC at $T<T_{c}$ is a 
direct manifestation of the relativistic low-energy Dirac fermions in silicene.
 
\begin{figure}[!thpb]
\centering
\includegraphics[width=0.99\linewidth]{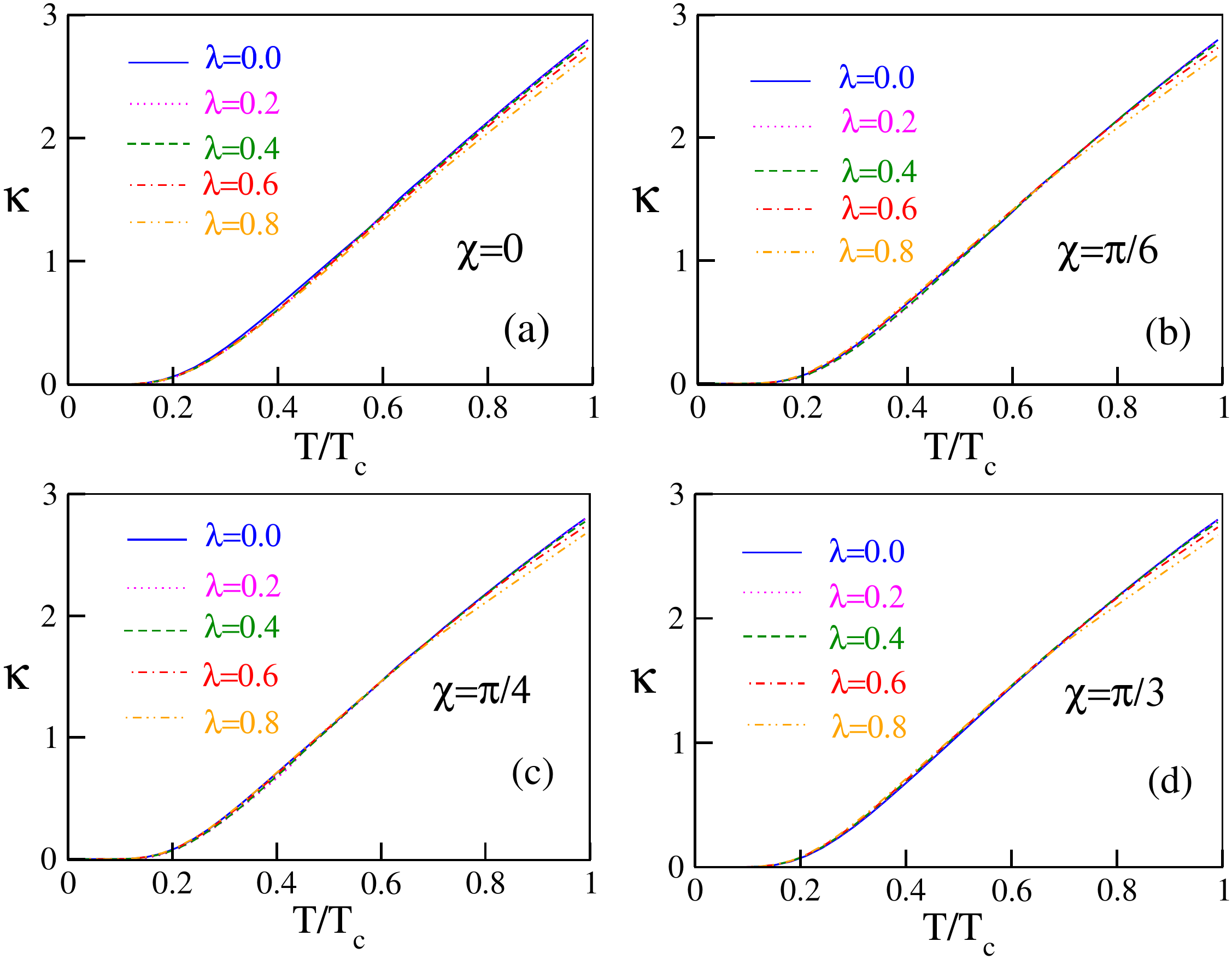}
\caption{(Color online) Thermal conductance is shown as a function of temperature $T/T_C$ with $U=100\Delta$ and $\lambda$ ranging 
from 0 to 0.8 for moderate doping ($\mu_{N}=0.5\Delta$). 
}
\label{KTMu5}
\end{figure}

\begin{figure}[!thpb]
\centering
\includegraphics[width=1.0\linewidth]{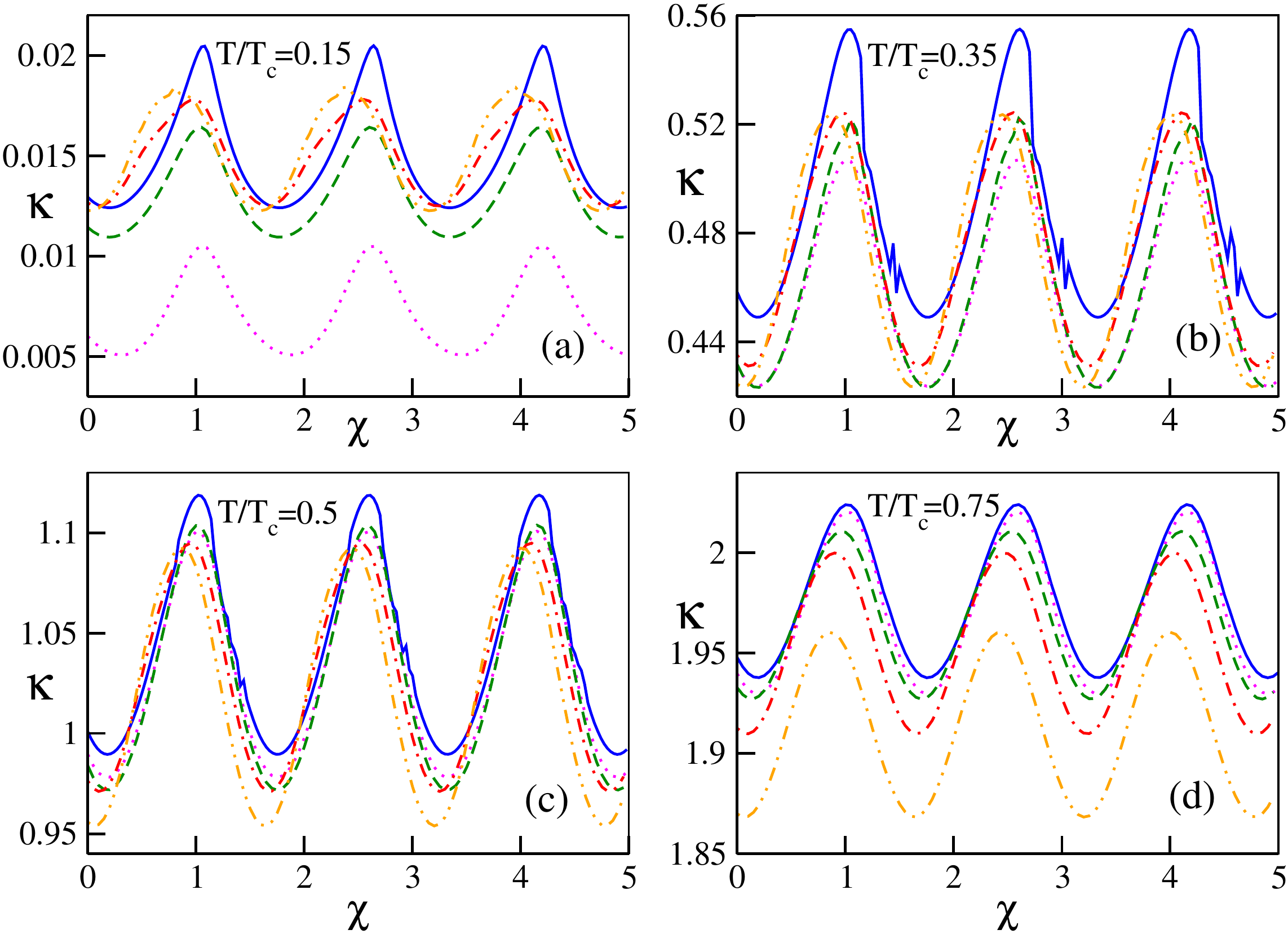}
\caption{(Color online) We show the thermal conductance as a function of the barrier strength $\chi$ with $U=100\Delta$ and $\lambda$ 
ranging from 0 to 0.8 for moderate doping ($\mu_{N}=0.5\Delta$). 
Specification of $\lambda$ is same as in Fig.~\ref{KChiMu0}.}
\label{KChiMu5}
\end{figure}

\subsection{Moderately doped regime ($\mu_{N}\neq 0$)}
In this subsection, we present our results for the moderate doping case where chemical potential in the normal part of the silicene sheet 
is $0.5\Delta$. This regime is qualitatively different from the undoped one because the doping concentration has now almost same order of 
magnitude with $\lambda$. So it is interesting to analyse whether non-trivial behavior of TC emerges out due to the interplay between 
doping and band gap at the two valleys. In Fig.~\ref{KTMu5}((a)-(d)), TC is shown as a function of temperature with different $\lambda$ and
for various barrier strength $\chi$. The striking difference from the undoped case is that $\kappa$ does not show monotonic behavior with $\lambda$. 
When $T\ll T_{c}$, $\kappa$ decreases with increasing $\lambda$ value by $E_{z}$ from 0 to 0.4. Then $\kappa$ further increases as $\lambda$ 
crosses $\mu_N$ value. At temperaure close to $T_c$, $\kappa$ decreases monotonically with increasing $\lambda$ similar to the undoped case.
Note that, $\kappa$ decreases in the $T\ll T_{c}$ regime due to the evanescent modes present between the energy range $|\mu_{N} - \lambda|$ 
to $|\mu_{N} + \lambda|$. Then $\kappa$ start increasing in the subgapped regime when $\mu_{N} \sim \lambda$ resulting in the non-monotonic
behavior. As long as $T \rightarrow T_{c}$ it again decreases due to the silicene band gap like the $\mu_{N}=0$ case. 

Transition from non-monotonic to monotonic behavior of TC takes place at $T\,\sim\,0.6 T_c$ independent of $\chi$ value. This non-monotonic 
characteristics is more promiment in Fig.~\ref{KChiMu5}((a)-(b)) where oscillatory nature of $\kappa$ with barrier strength is presented for 
different values of $T/T_c$. For a fixed $T/T_{c}$, such non-monotonic characteristics of $\kappa$ can be tuned by the external electric 
field $E_{z}$ which is unique in silicene. Here also the periodicity of oscillations remains $\pi/2$ independent of temperature and contribution 
in $\kappa$ originates from both SAR and retro AR. 

\subsection{Highly doped regime ($\mu_{N}\sim 100\Delta$)}
\begin{figure}[!thpb]
\centering
\includegraphics[width=0.99\linewidth]{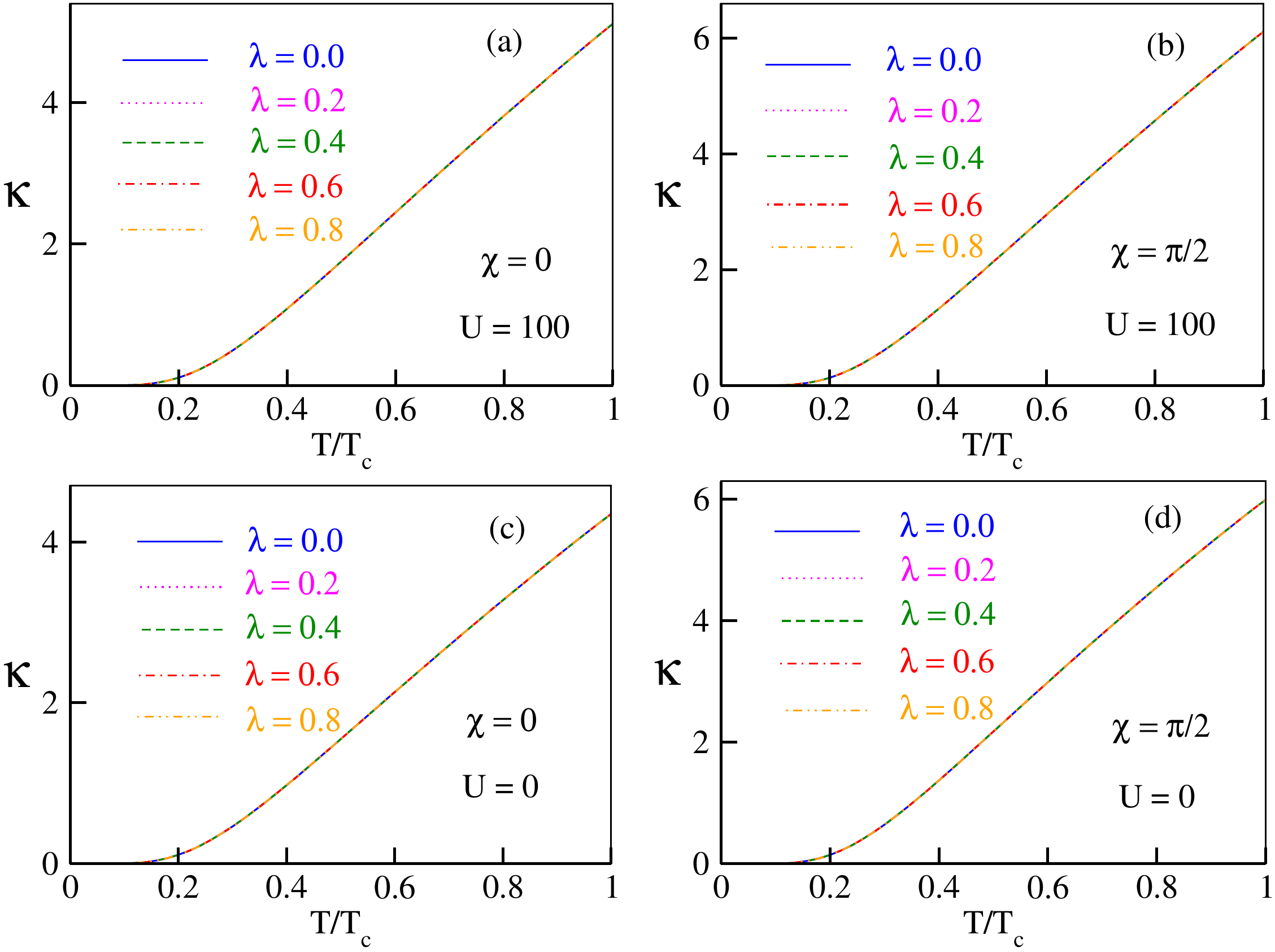}
\caption{(Color online) Thermal conductance is shown as a function of temperature $T/T_C$ with $\lambda$ ranging from 0 to 0.8 
and $\lambda^{\prime}$ ranging from 40 to 40.8 for the highly doped ($\mu_{N}\sim100\Delta$) regime. 
}
\label{KT}
\end{figure}

\begin{figure}[!thpb]
\centering
\includegraphics[width=1.0\linewidth]{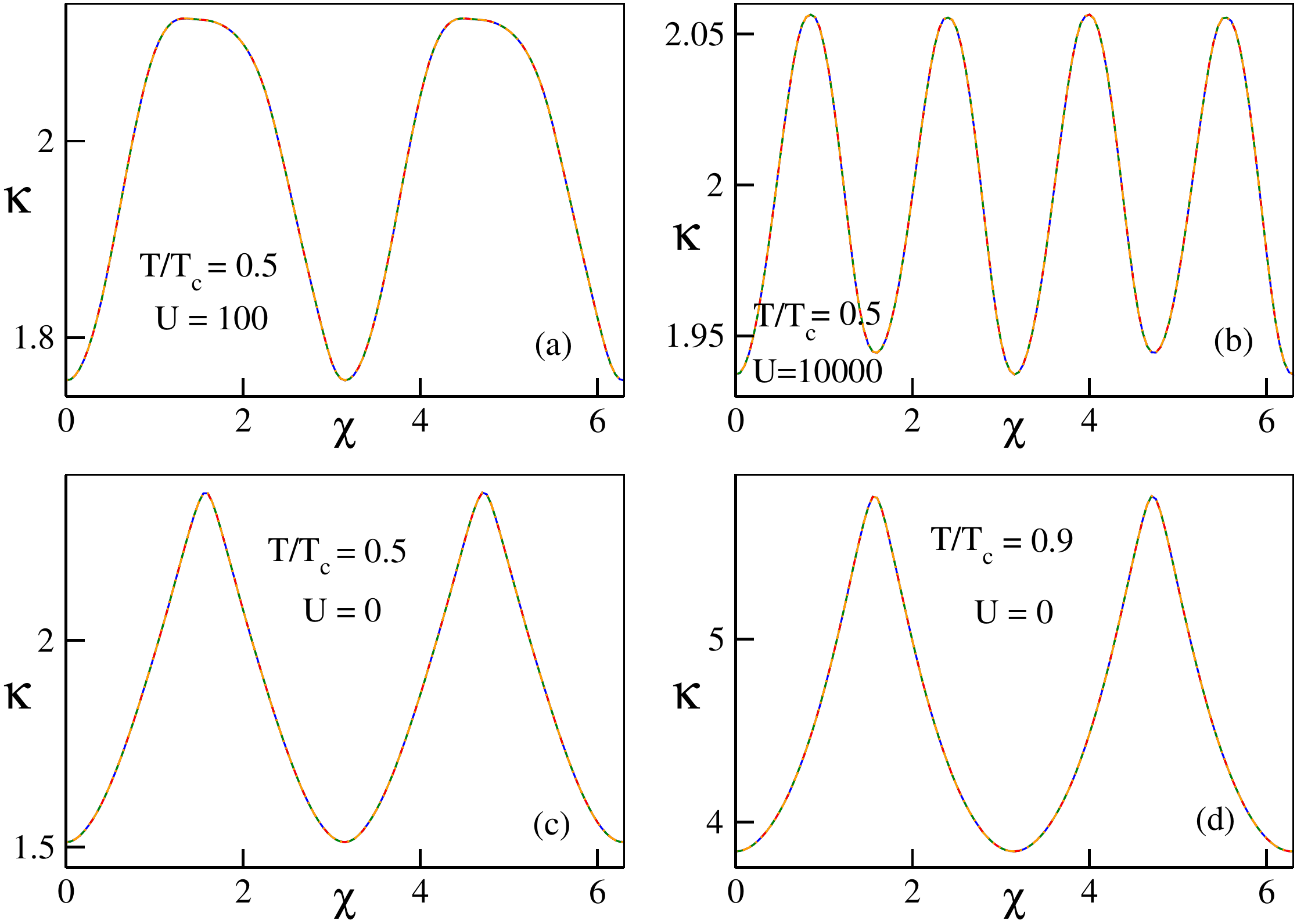}
\caption{(Color online) Thermal conductance is shown as a function of the barrier strength $\chi$ with $\lambda$ ranging from 0 to 0.8 for high 
doping ($\mu_{N}\sim100\Delta$) condition. 
Specification for $\lambda^{\prime}$ remains same as mentioned in Fig.~\ref{KT}.}
\label{KChi}
\end{figure}

Here in this subsection we present the features of TC while normal portion of silicene is highly doped ($\mu_N\,\sim\,100\Delta$). 
In this case the mean-field condition: $\mu_{N}+U\gg \Delta$~\cite{beenakker1} can be satisfied by assuming $U\ll\Delta$ or taking $U\gg\Delta$ 
as before. Former one does not exhibit any Fermi surface mismatch between the normal and superconducting regions. On the other hand, the latter one
contributes to large density mismatch between the two sides. We have numerically calculated $\kappa$ for $U=0\,\ll\,\mu_N$, 
$U=100 \Delta \simeq \mu_N$ and $U=10000\Delta\gg\mu_N$ regime. The corresponding results are presented in Fig.~\ref{KT} and Fig.~\ref{KChi}. 
Here also, similar to the undoped and moderately doped regime, $\kappa$ has exponential dependance on temperature which is a universal 
feature in thermal transport. The only difference from the previous two cases lies in the fact that we consider the separate contribution of 
both the valleys ${\bf {K}}$ and ${\bf {K}^{\prime}}$ when $\mu_N \gg \Delta$ (see Fig.~\ref{band}). 

From analytical expressions of the superconducting wave functions (see Eq.(\ref{sc1}) and Eq.(\ref{sc2})), we notice that the change in wave 
functions due to the variation of $\lambda$ and $\lambda^{\prime}$ is negligible because $\mu_{N}\sim 100\Delta \gg \lambda,\lambda^{\prime}$. 
Hence, in this regime $\kappa$ comes out to be independent of the applied electric field $E_{z}$ which is depicted in Fig.~\ref{KT}. 
The corresponding behavior is independent of $U$ also. Nevertheless, the quantitative value of $\kappa$ is enhanced by a factor 
of ``$2$'' compared to the previous two cases due to the contribution coming from both the valleys. 

The oscillatory behavior of TC with respect to the barrier strength $\chi$ persists in the highly doped regime as well (see Fig.~\ref{KChi}((a)-(d))). 
However, now the periodicity changes with the $U$ value. As long as $U\gg\mu_N$, period remains $\pi/2$ but it increases gradually to $\pi$ 
as $U$ decreases towards $U\ll\mu_N$. Both for $U=0$ and $U=100\Delta$, periodicity of $\kappa$ remains same at $\pi$ but the spread of the 
curve decreases as $U$ decreases as depicted in Fig.~\ref{KChi}(a,c). This change of behavior with variation of $U$ can be qualitatively understood 
from Fermi surface mismatch between the normal and superconducting sides. For large Fermi wavelengths mismatch between the normal and superconducting 
regions, period of oscillations remains $\pi/2$ which is similar to the undoped and moderately doped regimes. However, as the Fermi wavelengths 
mismatch becomes vanishingly small in the highly doped regime, periodicity of oscillation converts to $\pi$. Here also, $\lambda$ as well as
$\lambda^{\prime}$ have neglizible effect on the thermal transport as $\mu_N$ is the dominant energy scale in this particular regime. Similar 
periodicity of $\pi$ in the behavior of tunneling conductance in graphene for the highly doped regime was reported earlier in Ref.~\onlinecite{subhro}.

Note that, for the highly doped regime, major contribution in $\kappa$ originates from the retro AR in contrast to SAR in the undoped regime. 
Also the periodicity of $\kappa$ changes from $\pi/2$ to $\pi$ as long as $U\approxeq\mu_N$. Such change of periodicity with doping, in the 
behavior of thermal conductance in the thin barrier limit, can be an indirect way to identify the crossover from SAR to retro AR in Dirac materials.
Although, it is not apparent to compute separately the individual contribution of retro AR and SAR to $\kappa$ when $\mu_{N}\neq0$. 
This is because within our scattering matrix formalism we have to average over all values of energy (see Eq.(\ref{kappa})). Hence, the change of 
periodicity of $\kappa$ from $\pi/2$ to $\pi$ may not be a strong justification (smoking gun signal) for the crossover phenomenon from SAR to 
retro AR as the periodicity again can change from $\pi$ to $\pi/2$ due to Fermi wavelengths mismatch between the normal and superconducting regions 
even if $\mu_{N}\sim100\Delta$ where the major contribution to $\kappa$ arising from retro AR (see Fig.~\ref{KChi}(b)). 
However, to observe the latter change, one has to enhance the doping concentration in the superconducting side also.

\section{Thick barrier}  {\label{sec:IV}}
In this section we examine TC in the thick barrier limit where we consider a barrier of width $d$ and height $V_0$. The height of the barrier can
be tuned by applying an additional gate voltage in the insulating region~\cite{bhattacharjee2007theory}. We emphasize on the role being played 
by the barrier height $V_0$ as well as thickness $d$. We show $\kappa$ manifests osscillatory behavior with respect to both $d$ and $V_0$. 
However, the period of oscillation is no longer universal as in the thin barrier limit but beocmes a function of applied voltage 
$V_0$ and width $d$. Similar feature is found earlier in graphene NIS junction ~\cite{bhattacharjee2007theory} where tunneling conductance 
is shown to have oscillation whose period depends on $V_0$. 

 Note that, in the thick barrier limit, extended BTK formalism~\cite{blonder1982transition} is valid for our model of NIS junction if 
$d\leq \xi$ where $\xi=\hbar{v_F}/\pi\Delta$ which is the phase coherence length in the superconducting side. Fermi wavelength is given by, 
$\lambda_{F}=2\pi/k_F$ where ${k_F}\,=\mu_N/\hbar v_F$ being the Fermi wave vector. So $\lambda_{F}$ and $\xi$ are 
related by, \,$\lambda_{F}=2\pi^2 \Delta~\xi/\mu_N$. We notice that undoped regime is not valid in the thick barrier limit becuase 
Fermi wavelength diverges in that regime. In the moderately doped regime, choosing $\mu_N=0.5 \Delta$ as before, we obtain 
$d/\lambda_{F}\leq 1/4\pi^2 \sim 0.025$. 

\vspace {-0.54cm}
\subsection{Moderately doped regime ($\mu_{N}\neq 0$)}
\begin{figure}[!thpb]
\centering
\includegraphics[width=1.0\linewidth]{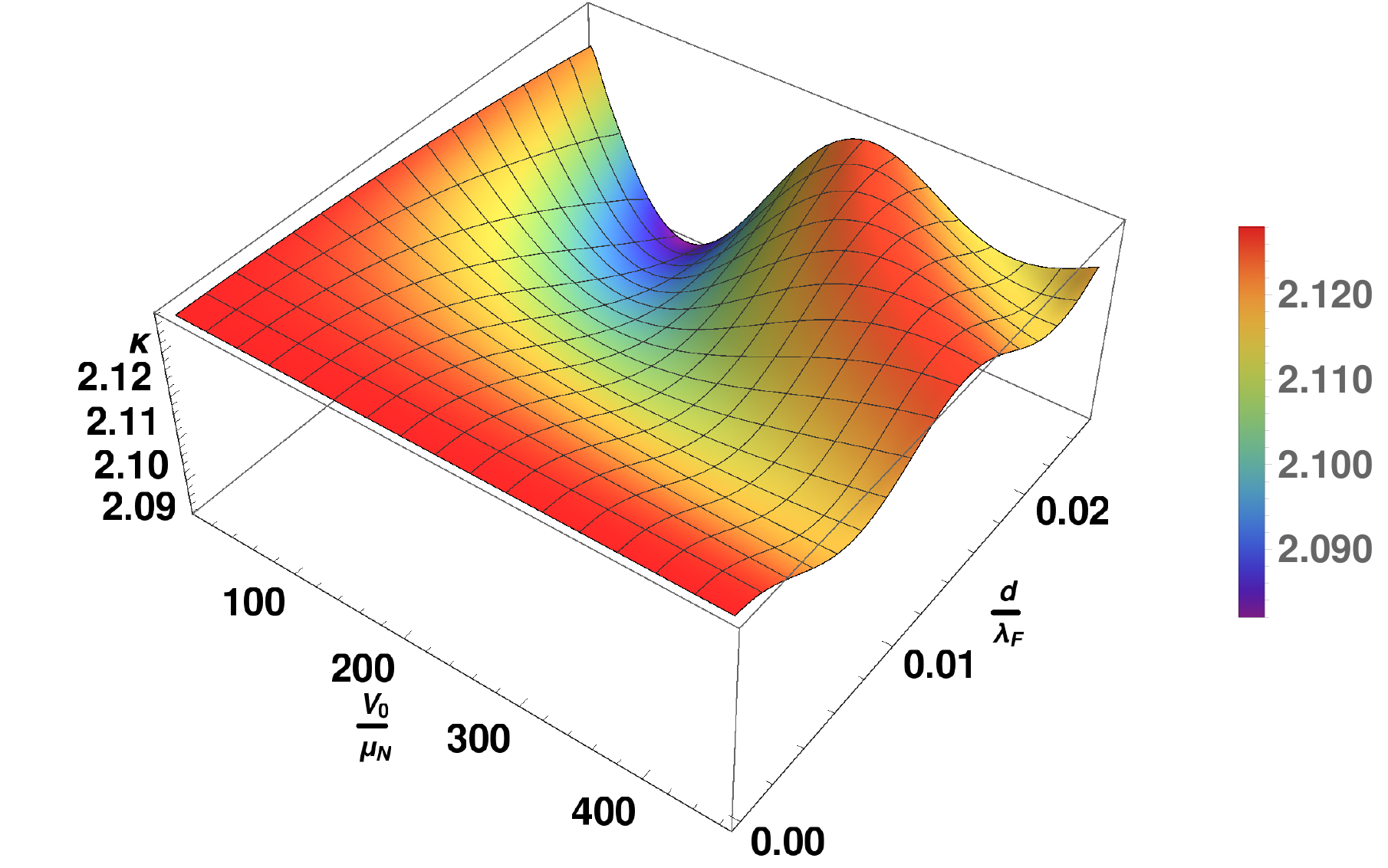}
\caption{(Color online) Plot of thermal conductance as a function of the barrier height $V_0$ and barrier thickness $d$ for $T/T_c=0.8$, 
$\lambda=0.3$, $U=100\Delta$ and $\mu_{N}=0.5\Delta$.}
\label{thick1}
\end{figure}

\begin{figure}[!thpb]
\centering
\includegraphics[width=1.0\linewidth]{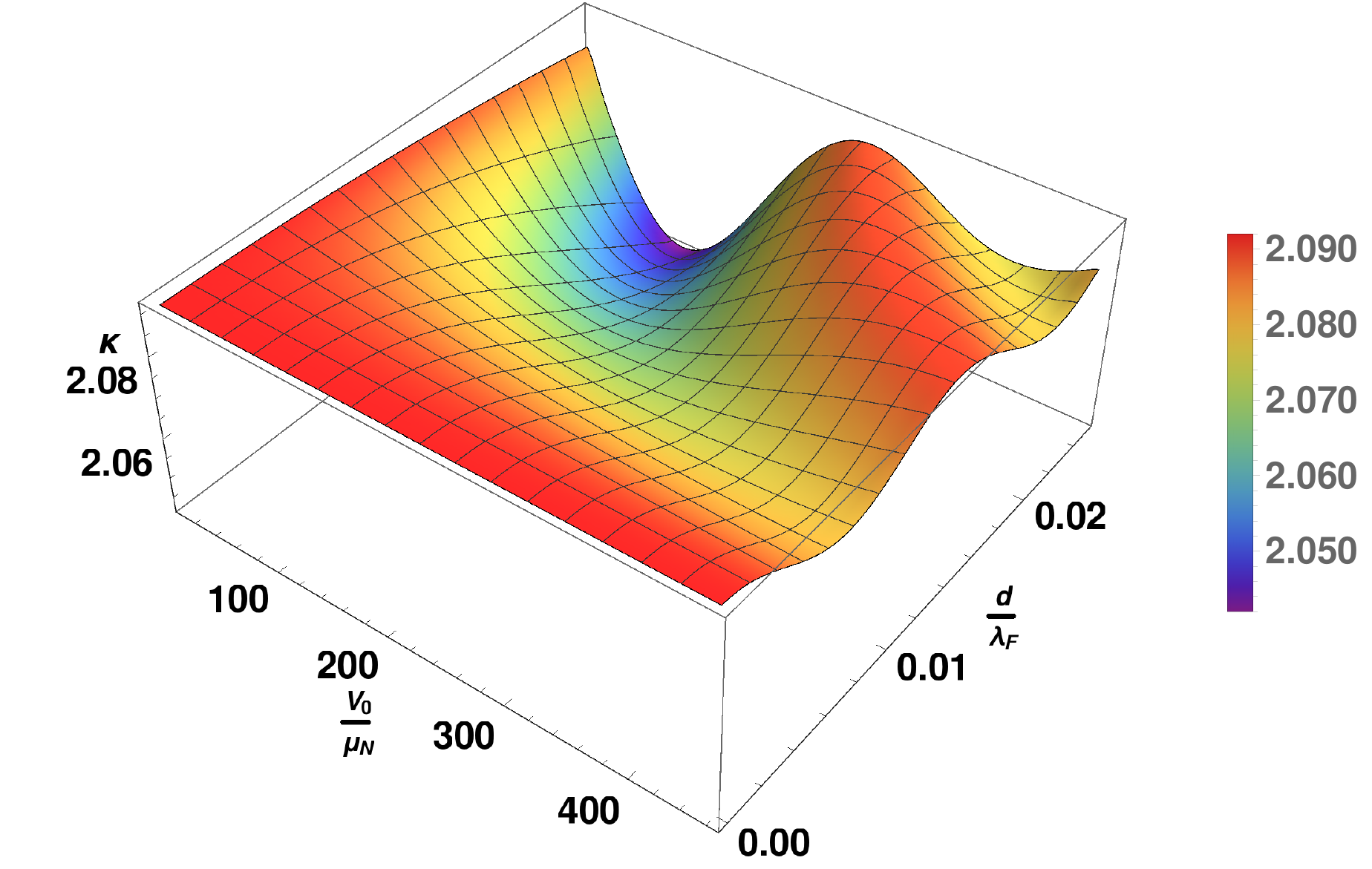}
\caption{(Color online) Thermal conductance is shown as a function of the barrier height $V_0$ and barrier thickness $d$. Here $\lambda=0.7$ 
and the value of the other parameters are chosen to be the same as in Fig.~\ref{thick2}.
}
\label{thick2}
\end{figure}

When the doping concentration is moderate ($\mu_{N}=0.5\Delta$) in the normal silicene regime, TC exhibits similar features as in the 
thin barrier limit. Here we illustrate the behavior of TC as a function of barrier height $V_0$ and thickness $d$ in Fig.~\ref{thick1} 
and Fig.~\ref{thick2} for $\lambda=0.3$ and $\lambda=0.7$ respectively. We note the following features. (i)~When $d \rightarrow 0$, 
TC is unaffected by the barrier height $V_{0}$. This is true for arbitray bandgap $\lambda$ as we can see from Fig.~\ref{thick1} 
and Fig.~\ref{thick2}. Nonetheless, $V_{0}$ affects TC as $d$ increases. Qualitatively we understand that as $U$ is chosen to be 
large $\sim\,100\Delta$, small barrier height $V_{0}$ has negligible effect on TC. 
(ii)~As barrier height dominates $U$, TC exhibits oscillatory behavior as a function of $d$ and such oscillation persists even for very
large values of $V_{0}$. 
Similarly oscillation is present as $V_0$ changes even for $d \sim 0.025 \lambda_{F}$. However, the period of oscillation 
does not show any universal periodicity of $\pi/2$ like in the thin barrier case. The period of oscillation of $\kappa$ depends on 
both $d$ and $V_0$. Similar feature was found earlier in case of tunneling conductance in graphene NIS junction~\cite{bhattacharjee2007theory}. 
(iii)~The external electric field $E_{z}$ does not change the qualitative behavior of $\kappa$ as shown in Fig.~\ref{thick1} 
and Fig.~\ref{thick2}. Although it changes the quantitative value of $\kappa$. As $\lambda$ increases by tuning $E_{z}$, TC reduces 
monotonically with both $d$ and $V_0$ similar to the thin barrier case when $T/T_{c}=0.8$.

\subsection{Highly doped regime ($\mu_{N}\sim 100$)}
Here, we present the behavior of TC as a function of $d$ and $V_0$ with high doping concentration where $\mu_{N}\sim 100\Delta$. 
We choose $U=0$ only. Hence, there is no Fermi wavelength mismatch between the normal and superconducting side of the silicene sheet. 
Thus the effect of applied gate voltage $V_{0}$ across the insulating region can be investigated prominently in this regime due to $U=0$. 
Also, as we have already pointed out in thin barrier limit that $\lambda$ and $\lambda^{\prime}$ has negligible effect on $\kappa$ when 
$\mu_N/\Delta\gg \lambda,\lambda^{\prime}$, hence we consider $\lambda=0$ and $\lambda^{\prime}=40$. 

\begin{figure}[!thpb]
\centering
\includegraphics[width=1.0\linewidth]{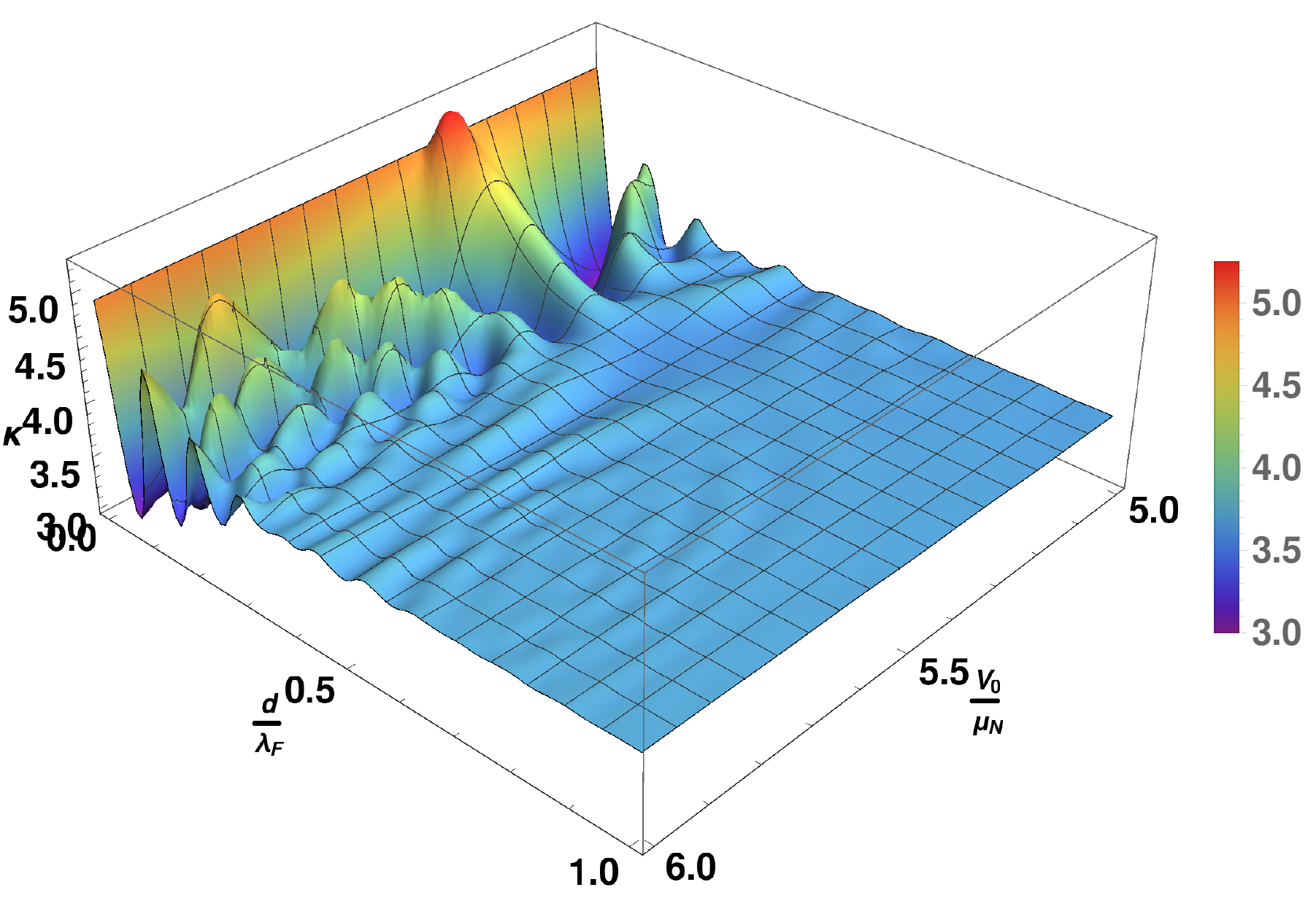}
\caption{(Color online) Thermal conductance is depicted as a function of barrier height $V_0$ and thickness $d$ with $\lambda=0$, 
$\lambda^{\prime}=40$, $U=0$ and $T/T_c=0.8$ for the highly doped ($\mu_{N}\sim100\Delta$) regime. 
}
\label{thick3}
\end{figure}

Fig.~\ref{thick3} represents TC as a function of $d$ and $V_0$ for $\lambda=0$ and $T/T_c=0.8$. We choose $V_0$ value to be much larger 
than $\mu_N$ in order to investigate the effect of applied gate voltage or barrier height on TC. We note that $\kappa$ exhibits oscillation 
with respect to $V_{0}$ even for very small barrier thickness $d$. The period of these oscillations is entirely dependent on $V_0$. 
As mentioned earlier, such oscillations of $\kappa$ at very small $d$ does not appear at moderate doping concentration unless and until
$V_{0}$ exceeds $U$. Note that, the enhancement in the quantitative value of $\kappa$ compared to the previous case arises due to both 
${\bf {K}}$ and ${\bf {K}^{\prime}}$ valley contribution. 
Also in the highly doped regime, the amplitudes of oscillations of $\kappa$ decay after a certain value of 
barrier thickness ($d\sim 0.4\lambda_{F}$) for arbitrary barrier height $V_{0}$. This can be understood from the Fermi wave-length
mismatch between the barrier and the normal silicene region for high value of $d$ and $V_{0}$.
This feature of TC is in sharp contrast to the tunneling conductance in graphene which is oscillatory for arbitrary 
$d$ and $V_{0}$~\cite{bhattacharjee2007theory}.

\section{Summary and conclusions} {\label{sec:V}}
To summerize, in this article, we investigate thermal conductance $\kappa$ by Dirac fermions in silicene NIS junction where superconductivity 
is induced in silciene sheet through the proximity effect. We study the behavior of TC in this set-up both for thin and thick insulating barrier 
limit. We show that TC exhibits $\pi/2$ periodic oscillation with respect to the barrier strength in thin barrier limit for undoped ($\mu_N=0$) 
and moderately doped ($0<\mu_N\leq \Delta$) regime where the Fermi surface mismatch between the normal and superconducting sides is significant. 
The oscillation becomes $\pi$ periodic as a function of barrier strength in the highly doped ($\mu_{N}\gg\Delta$) regime where Fermi surfaces 
in the two sides are almost aligned. This change of periodicity ($\pi/2$ to $\pi$) in thermal response with the variation of doping concentration 
can be an indirect probe to identify the crossover from SAR to retro AR. Nonetheless, TC shows conventional exponential dependence on temperature 
independent of doping concentration and barrier characteristics. The external electric field reduces TC monotonically in the undoped regime. 
However, a non-trivial interplay between band gap at Dirac points and doping concentration appears in the moderately doped case. Consequently, 
electric field can tune TC in the later regime. On the other hand, electric field has negligible effect on TC when $\mu_N/\Delta\gg \lambda$.

In the thick barrier limit, oscillation of TC persists both as a fucntion of barrier thickness $d$ as well as barrier height $V_{0}$.
The latter can be tuned by an additional gate voltage appled at the insulating region. However, we show that the periodicity of TC
no longer remains constant, rather becomes functions of both $d$ and $V_{0}$. Also after a certain barrier thickness ($d\sim 0.4\lambda_{F}$), 
amplitude of oscillations in TC decays for arbitrary $V_{0}$ in the highly doped regime.

In our analysis, we consider only the electronic contribution in TC and neglect the phonon contribution at small temperatures ($T<T_{c}$).
Very recently, nanoscale control of phonon excitation in graphene has been reported~\cite{kim2015nanoscale}. Hence, such nanoscale control 
of phonon excitation in silicene and the effect of electron-phonon interaction on TC will be presented elsewhere.

As far as experimental realization of our silicene NIS set up is concerned, superconductivity in silicene can be induced by $s$-wave
superconductor like $\rm Al$ ~\cite{heersche2007bipolar,choi2013complete}.
In recent years, proximity induced superconductivity has been observed in other 2D materials 
such as graphene~\cite{heersche2007bipolar,choi2013complete,calado2015ballistic} and transition metal dichalcogenides~\cite{shi2015superconductivity}. 
Once such superconducting proximity effect is realized in silicene, fabrication of silicene NIS junction can be feasible. 
Typical spin-orbit energy in silicene is $\lambda_{\rm SO}\sim 4~\rm meV$ while the buckling parameter $l\approx 0.23~\rm\AA$~\cite{liu2011low,ezawa2015}. 
Considering Ref.~\onlinecite{heersche2007bipolar}, typical induced superconducting gap in silicene would be $\sim 0.2~\rm meV$. 
For such induced gap, choosing $\mu_N\,\sim\,100\Delta\,\sim\,20~\rm meV$, we obtain $\lambda_{F}\,\sim\,130~\rm nm$. Hence, a barrier of thickness 
$\sim 10-15~\rm nm$ may be considered as thin barrier and the gate voltage $V_{0}\sim 500~\rm meV$ can therefore meet the demands of our 
silicene NIS setup. For the thick barrier limit, thickness can be varied arbitrarily (satisfying $d\,\leq\,\lambda_{F}\sim 100~\rm nm$), 
with the gate voltage $V_{0}\sim\,100-200~\rm meV$.

However, the effects of external electric field might not be visible in the above regime as envisaged by our theoretical calculation. 
To realize non-trivial effects due to the electric field on TC, chemical potential in the normal silicene region can be $\mu_N\sim 80-120~\rm \mu eV$
and the external electric field $E_{z}$ can be within the range $E_{z}\sim 170-180~\rm eV/\mu m$. 
In this moderately doped regime ($0<\mu_N\leq\Delta$), the criterion for $d$ and $V_{0}$ can be similar to the highly doped regime as 
mentioned before. 

Note that, in our analysis, we have considered a bulk silicene material following Ref.~\onlinecite{Linder2014}. 
The bulk-boundary correspondence has not been taken into account within our scattering matrix formalism. So, we cannot distinguish between 
the topological phase or the band insulating phase within our formalism even if we tune the electric field $E_{z}$ in our calculation. 
Hence, in our analysis, the contribution in the thermal conductance is arising from the bulk states only.

We expect our results to be analogous to the recently discovered 2D materials like germenene, stanene~\cite{davila2014germanene,zhu2015epitaxial}. 
Although the effect of Rashba SOC $\lambda_{R}$ in these materials can be more important than silicene~\cite{liu2011low,ezawa2015}. 
For silicene, $\lambda_R$ is small compared to $\lambda_{SO}$~\cite{ezawa2012topological}. The low energy spectrum of silicene is independent 
of $\lambda_{R}$ only at the Dirac point~\cite{ezawa2012topological}. Inclusion of small $\lambda_R$ breaks
the spin symmetry and spin is no longer a good quantum number. Qualitatively, from scattering point of view, presence of small $\lambda_R$ introduces 
spin flip scattering processes from the normal-superconductor (NS) interface. Apart from spin conserving reflection and AR processes, 
the reflection and AR processes with spin flip also contribute to $\kappa$. Nevertheless, as $\lambda_R$ is small, the amplitudes 
of those additional scattering processes will also be small. Hence, after averaging over all the energy values while computing $\kappa$, 
the contribution arising from these two extra scattering processes on the resulting thermal conductance will be vanishingly small. 
Thus the qualitative feature of $\kappa$ as a function of $T/T_{c}$ or $\chi$ will remain similar even one includes small $\lambda_R$ into account.

\acknowledgments{We acknowledge S. D. Mahanti for valuable discussions and encouragement at the initial stage of this work.}

\vspace{1cm}

\bibliography{Silicene_NIS_ref} 

\end{document}